\documentclass[3p,times,procedia]{elsarticle}
\usepackage{nupha_ecrc}
\usepackage[utf8]{inputenc}

\volume{00}

\firstpage{1}

\journalname{Nuclear Physics A}

\runauth{Gardim, Grassi, Luzum, Noronha-Hostler }

\jid{nupha}

\jnltitlelogo{Nuclear Physics A}

\usepackage{amssymb}

\usepackage[figuresright]{rotating}

\begin{document}

\begin{frontmatter}



\dochead{XXVIth International Conference on Ultrarelativistic Nucleus-Nucleus Collisions\\ (Quark Matter 2017)}

\title{Mixed Harmonic Correlations: Hydrodynamics Predictions at RHIC using Experimental Analysis Techniques }


\author{Fernando G. Gardim\corref{blue}\fnref{a}}
\ead{fernando.gardim@unifal-mg.edu.br}
\author[b]{Frederique Grassi}
\author[b]{Matthew Luzum}
\author[c]{Jacquelyn Noronha-Hostler}
\address[a]{Instituto de Ci\^encia e Tecnologia, Universidade Federal de Alfenas, Cidade Universit\'aria, 37715-400 Po\c cos de Caldas, MG, Brazil}
\address[b]{Instituto de F\'{i}sica, Universidade de S\~{a}o Paulo, 05315-970 S\~{a}o Paulo, SP, Brazil}
\address[c]{Department of Physics, University of Houston, Houston, TX 77204, USA}
\begin{abstract}

Correlations of different azimuthal flow harmonics $v_n$ and symmetry planes $\Psi_n$ can add constraints to theoretical models, and probe aspects of the system that are independent of the traditional single-harmonic measurements. Using NeXSPheRIO, a hydrodynamical model which has accurately reproduced a large set of single-harmonic correlations at RHIC, we make predictions of these new observables for 200 A GeV Au+Au Collisions, providing an important baseline for comparison to correlations of flow harmonics, which contain non-trivial information about the initial state.  We also point out how to properly compare theoretical calculations to measurements using wide centrality bins and non-trivial event weighting.

\end{abstract}


\end{frontmatter}


\section{Introduction}
\label{}

Ultrarelativistic nuclear collisions produce the Quark Gluon Plasma (QGP) and after a short period of time  equilibrium is reached. The system thus expands and cools as a relativistic fluid until a certain freeze-out temperature is reached after which particles are emitted. Theoretical studies of the QGP require an initial condition of the energy density profile immediately after the collision, which is then evolved solving relativistic hydrodynamics equations. As the initial state of the QGP is poorly constrained at present, it is important to obtain as many details as possible of the final distribution of particles, whereas initial anisotropy (coordinate space) defines unequivocally collective anisotropic flow (momentum space), thus probing indirectly the initial anisotropy. In theory, the emitted particles distribution can be written as a Fourier series: 

\begin{equation}
P(\phi) = \frac 1 {2\pi} \sum_n V_n e^{-in\phi},
\end{equation}
where $V_n\equiv v_n e^{in\Psi_n}$, is the flow vector, with $v_n$ and $\Psi_n$ as the magnitude and orientation of $V_n$. These flow vectors fluctuate significantly from one event to the next, even within a particular centrality class. Thus, there is not a small set of constant coefficients $\{v_n, \Psi_n\}$, but instead a large set of statistical properties. When one considers the alignment and correlation between flow vectors of different harmonics, it opens up a door to a large number of additional measurements. For instance, it provides mechanisms to isolate linear and non-linear hydrodynamic response  \cite{Giacalone:2016afq}, it sheds light on the temperature dependence of $\eta/s(T)$  \cite{Niemi:2015qia}, and it allows one to obtain information about flow vs. non-flow effects from data. The tool able to do this is called mixed-harmonic correlations.

Many mixed-harmonic correlations have now been performed by LHC collaborations \cite{ALICE:2016kpq}, but have not yet been done at RHIC, leaving an opportunity to make predictions.

In this work we present predictions for some mixed-harmonic correlations at RHIC $\sqrt{s_{NN}}=200$ GeV, and also present details of experimental analyses, such as centrality binning and event weighting, which can have important effect in the comparison with data.

\section{Mixed Harmonic Correlation}
\label{mhc}

When only collective flow is present, particles are emitted independently, and the distribution of particles for $n$ observables $\phi_1$, $\phi_2$, ... $\phi_n$, factorizes:

\begin{equation}
P(\phi_1,\ldots,\phi_n)\stackrel{\rm{(flow)}}{=}P(\phi_1)\ldots P(\phi_n). 
\end{equation}

Using this property for the basic building block of correlation measurements, the general $m-$particle correlator \cite{Luzum:2013yya} can be written as

\begin{equation}
\langle m \rangle_{n_1, n_2, \ldots, n_m}\stackrel{\rm{(flow)}}{=} \left\langle v_{|n_1|}^{a_1} v_{|n_2|}^{a_2} \ldots v_{|n_m|}^{a_m} \cos(n_1\Psi_{n_1}^{a_1} + n_2\Psi_{n_2}^{a_2}\ldots n_m \Psi_{n_m}^{a_m}) \right\rangle
\label{cumu}
\end{equation}
where $\langle \cdots \rangle \equiv \sum_{\rm events} W \ldots / \sum_{\rm events} W$, with $W$ as weight. In this work $W=M!/(M-m)!$, related with the number of charged hadrons in each event. Because each collision has a random azimuthal orientation, one can only measure rotation-invariant quantities. Thus, the non-zero correlators have $\sum n_i =0$ \cite{Gardim:2016nrr}.

The simplest measurement using the $m-$particle correlator is for $m=2$ known as the two-particle cumulant, $v_n\{2\}$. However, two particle correlations can have non-negligible nonflow contributions. Correlations between more than two particles diminish nonflow effects, since the correlation between a small  number of particles is suppressed. In order to suppress low-order nonflow correlations, one can measure the mixed harmonic observable called symmetric cumulant SC$(n,m)$, which is based on 4-particle correlations. Nevertheless, the information can be best viewed with a normalized version of the correlation, 
\begin{equation}
\label{NSC}
{\rm NSC}(n,m) \equiv \frac {{\rm SC}(n,m)} {\langle 2 \rangle_{n,-n}\langle 2 \rangle_{m,-m}} \stackrel{\rm{(flow)}}{=}  \frac {\langle v_n^2 v_m^2 \rangle -  \langle v_n^2\rangle \langle v_m^2 \rangle} {\langle v_n^2\rangle \langle v_m^2 \rangle}, \hspace{.6cm}     (n\neq m).
\end{equation}


%

These momentum-integrated measurements are only sensitive to the magnitude squared of the flow vector, $v_n^2$.  In order to gain information about the correlations of the entire momentum-integrated flow vector $V_n$, including its direction, one must consider other correlations, for instance, the event plane correlations:
\begin{eqnarray}
\label{EP24}
\langle \cos 4\left(\Phi_2 -  \Phi_4 \right) \rangle\{ {\rm SP}\} \equiv \frac { \langle 3 \rangle_{2,2,-4}} {\sqrt{\langle 4 \rangle_{2,2,-2,-2} \langle 2\rangle_{4,-4}}} \stackrel{\rm{(flow)}}{=} \frac {\langle v_2^2 v_4 \cos 4(\Psi_2 - \Psi_4) \rangle} {\sqrt{\langle v_2^4 \rangle \langle v_4^2\rangle}}\\
\label{EP23}
\langle \cos 6\left(\Phi_2 -  \Phi_3 \right) \rangle\{{\rm SP}\} \equiv \frac { \langle 5 \rangle_{2,2,2,-3, -3}} {\sqrt{\langle 6 \rangle_{2,2,2,-2,-2,-2} \langle 4\rangle_{3,3,-3,-3}}} \stackrel{\rm{(flow)}}{=} \frac {\langle v_2^3 v_3^2 \cos 6(\Psi_2 - \Psi_3)\rangle} {\sqrt{\langle v_2^6 \rangle \langle v_4^4\rangle}}\\
\label{EP235}
\langle \cos (2\Phi_2 +  3 \Phi_3  -  5\Phi_5 )  \rangle  \{{\rm SP}\} \equiv \frac { \langle 3 \rangle_{2,3,-5}} {\sqrt{\langle 2 \rangle_{2,-2} \langle 2 \rangle_{3,-3} \langle 2\rangle_{5,-5}}} \stackrel{\rm{(flow)}}{=} \frac {\langle v_2 v_3 v_5 \cos (2\Psi_2 + 3\Psi_3 - 5\Psi_5) \rangle} {\sqrt{\langle v_2^2 \rangle \langle v_3^2\rangle \langle v_5^2 \rangle}}.
\end{eqnarray}

For the calculations NeXSPheRIO \cite{Hama:2004rr} is used, which provides a reasonable description of Au+Au RHIC $\sqrt{s_{NN}}=200$ GeV data \cite{Gardim:2012yp}. Some of the quantities fitted are: rapidity and transverse momentum spectra, directed-elliptic flow, anisotropic flow, 2-particle correlation, trigger angle dependence, rapidity flow fluctuation. NeXus provides event-by-event initial conditions that includes initial flow and longitudinal fluctuations and SPheRIO solves ideal 3+1D hydrodynamic equations at finite $\mu_B$. We include results here for 6 centralities bin between 0-60\% and include only charged particles with $|\eta|<1$ (details of the code can be found in \cite{Gardim:2016nrr}).

It is known that large scale geometric properties of the initial conditions, called eccentricities $\varepsilon_n$, are related to the final $v_n$ \cite{Gardim:2011xv}: $v_2$ is understood as a response to the almond-shaped overlap area $\varepsilon_2$, $v_2\propto\varepsilon_2$; triangularity $\varepsilon_3$ is a very good predictor to $v_3$, $v_3\propto\varepsilon_3$; non-linear terms are necessary to predict $v_4$ and $v_5$ from initial energy density, $v_4\propto k\varepsilon_4+k'\varepsilon_2^2$ and $v_5\propto k\varepsilon_5+k'\varepsilon_2\varepsilon_3$ (these properties still hold even with bulk and shear viscosity\cite{Gardim:2014tya}). Thus, as $v_n\propto\varepsilon_n$ for $n\leq 3$, symmetric cumulants for $\varepsilon_n$ should be $\approx$ NSC(3,2),

\begin{equation}
\label{nsc}
\varepsilon SC(3,2) \equiv  \frac {\langle \varepsilon_3^2 \varepsilon_2^2 \rangle -  \langle \varepsilon_3^2\rangle \langle \varepsilon_2^2 \rangle} {\langle \varepsilon_3^2\rangle \langle \varepsilon_2^2 \rangle} \approx NSC(3,2).
\end{equation}

\begin{figure}
\centering
\includegraphics[width=15cm]{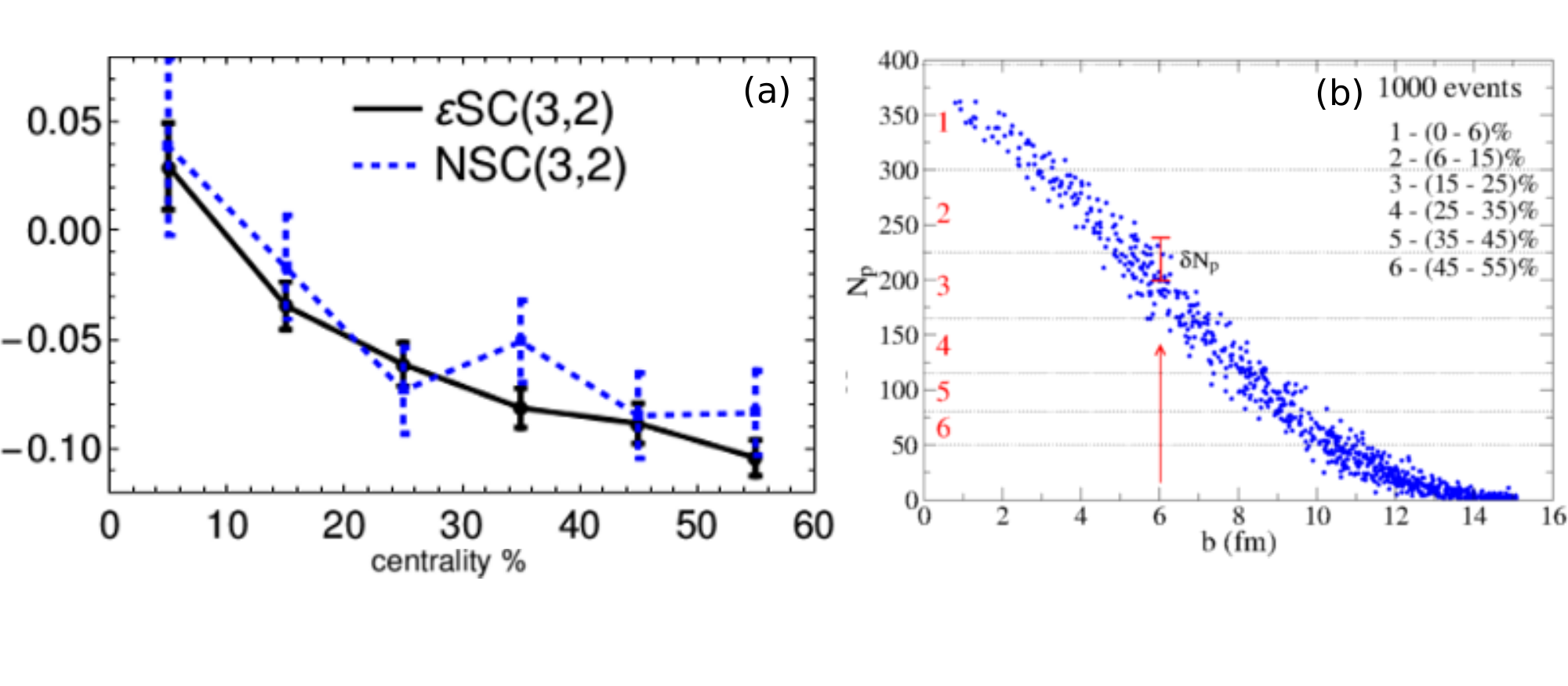}
\vspace{-0.8cm}
\caption{\label{vnen} (a) $\varepsilon{\rm SC}(3,2)$ and ${\rm NSC}(3,2)$ from NeXSPheRIO. Points are shifted horizontally for readability.  Errors bars obtained via jackknife resampling.(b) Correlation between impact parameter $b$ and number of participants $N_p$ using NeXSPheRIO. For central collisions the size of centrality entails large fluctuations.}
\end{figure}

This comparison is presented in  Fig. \ref{vnen} (a), where one can see a small difference between NSC(3,2) and $\varepsilon SC(3,2)$. The normalized symmetric cumulant for different models of initial conditions do not vary by a large amount \cite{Gardim:2016nrr}, even though these models are quite different. It is also important to point out the dependence of NSC(3,2) on collision energy is very small \cite{Gardim:2016nrr}. Thus one can study NSC(3,2) without the necessity of hydro events, implying it as a good observable to test models and parameters across collision energy.

\section{Predictions and Conclusions}

In order to make theory vs experiment comparisons, theorists much mimic experimental analysis, thus, experimental effects such as centrality rebinning and weights must be incorporated into theoretical calculations. 

The size of the centrality bin in the analysis can change measurements of $NSC(n, m)$, and this is mainly
due to the following effect: on average, more peripheral collisions have larger $v_n$, while more central collisions have smaller $v_n$. Thus for measurements using events in a large range of centrality, the impact parameter fluctuates significantly within the bin, Fig. \ref{vnen} (b). This trivial effect generates a spurious positive correlation \cite{Gardim:2016nrr}, compared to the value obtained when using narrow centrality bins. Since the dependence of $v_n$ on centrality is strongest in central collisions, this effect is most important there. While it is preferable to use small centrality bins, experimentally there may not be enough statistics to do so, which can be circumvented by first binning into small centrality bins and then recombining into larger bins i.e. ``centrality rebinning".

The other experimental procedure is the event-weight in the average computation. As LHC uses non-unity event-weights, we use the same weights for RHIC predictions (see Sec. \ref{mhc}). Additionally, these weights are also used in the centrality rebinning such that
\begin{eqnarray}
\label{recomb}
{\rm NSC}_{10\%}^{cen}&=&\frac{\sum_{c=1}^{10} {\rm NSC}_{1\%}^c \sum_{\rm events} W^c_{\langle m\rangle} }{\sum_{c=1}^{10} \sum_{\rm events}  W^c_{\langle m\rangle}},
\end{eqnarray}
where ${\rm NSC}_{1\%}^c$ is computed in a $c$ sub-bin for a centrality bin ($cen$). Throughout this work recombine with event weighing, Eq. \ref{recomb}, where a small centrality bin of 1\% will be recombined into 10\%, in order to obtain predictions to 200 A GeV Au-Au.

\begin{figure}
\hspace{-2cm}\includegraphics[width=18cm]{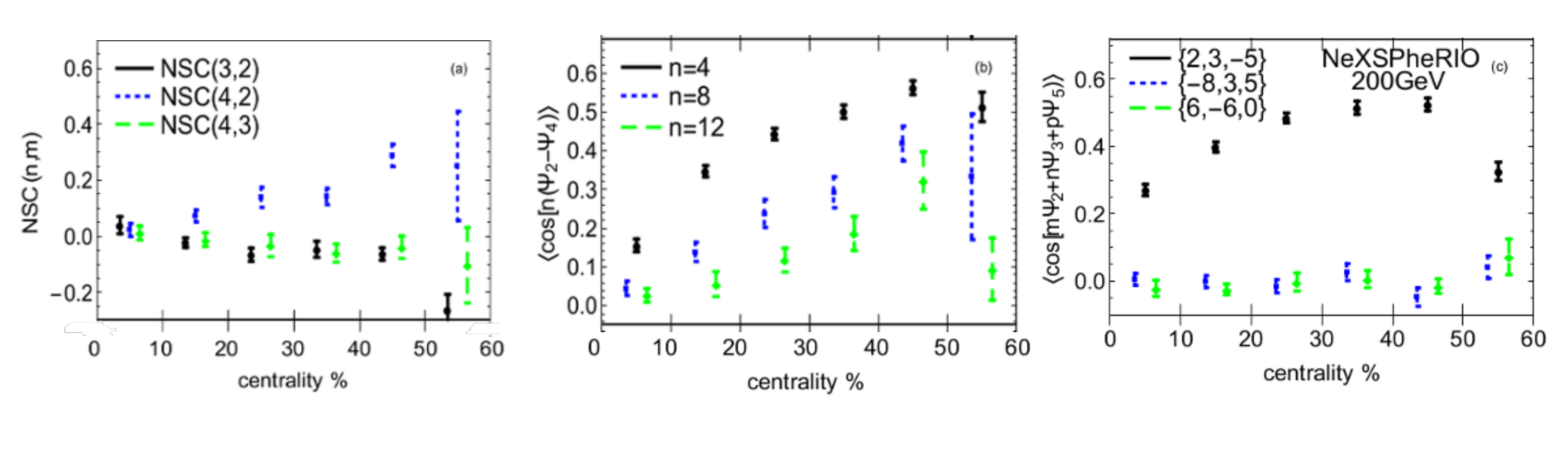}
\vspace{-0.8cm}
\caption{\label{pred} 200 A GeV Au+Au predictions from NeXSPheRIO for NSC(n, m) and mixed harmonic correlations, for $p_T>$ 200 MeV and $|\eta| < 1$. 
Error bars  obtained via jackknife resampling.}
\end{figure}

Predictions are presented in Fig. \ref{pred}. For symmetric cumulants Fig. \ref{pred} (a) one can note $v_2$ and $v_4$ are not correlated with $v_3$, but there is a strong correlation between $v_2$ and $v_4$ for peripheral collisions, due to non-linear effects \cite{Gardim:2011xv}, since $v_4$ can be mapped as response to the initial almond shape $\varepsilon_2$: $v_4\propto k\varepsilon_4+k'\varepsilon_2^2$. In the case of event plane correlations, Eqs. \ref{EP24}-\ref{EP235}, the correlation between $\Psi_2$ and $\Psi_4$ Fig. \ref{pred} (b), and between $\Psi_5$ and the combination of $\Psi_3$ and $\Psi_2$ Fig. \ref{pred} (c). 

To conclude, we have presented a precise quantitative prediction for RHIC's top energy, and the comparison with data will provide needed guidance to discriminate between different theoretical models. Our results are especially interesting as a baseline calculation since we have a model that is able to reproduce experimental with ideal hydrodyanmics and some of these observables are sensitive to $\eta/s(T)$. In addition, we showed that for correct predictions to data comparison, it is important to use centrality rebinning and event weighting. We also showed the results are in agreement with the expected non-linear response.\\

\textbf{Acknowledgments:} F.G.G. was supported by Conselho Nacional de Desenvolvimento Científico e Tecnologico (CNPq) no. 449694/2014-3, and Fundação de Amparo à Pesquisa de Minas Gerais FAPEMIG, no. APQ-02107-16. J.N.H. was supported by the National Science Foundation under grant no. PHY-1513864. F.G. was supported by Funda\c{c}\~ao de Amparo \`a Pesquisa do Estado de S\~ao Paulo no. 2015/00011-8 and 2016/03274-2, and CNPq no. 310737/2013-3.





\bibliographystyle{elsarticle-num}



\end{document}